\documentclass[3p, onecolumn]{elsarticle}

\usepackage{lineno,hyperref}
\modulolinenumbers[5]

\journal{arXiv}

\usepackage{amsmath}
\usepackage{amsthm}
\usepackage{subfigure}
\usepackage{color}
\usepackage{algorithm}
\usepackage{algpseudocode}
\usepackage{geometry}
\usepackage{graphicx}
\usepackage{lipsum}
\usepackage{balance}
\makeatletter

\newcommand{\ket}[1]{\left|#1\right\rangle}

\begin{document}

\begin{frontmatter}

\title{Quantum Anonymity for Quantum Networks}

\author{Awais Khan}

\author{Junaid ur Rehman}

\author{Hyundong Shin\corref{mycorrespondingauthor}}
\cortext[mycorrespondingauthor]{Corresponding author}
\ead{hshin@khu.ac.kr}

\address{Department of Electronic Engineering, Kyung Hee University, Yongin-si, 17104, Korea}

\begin{abstract}
We present the first quantum anonymous notification (QAN) protocol that introduces anonymity and paves the way for anonymous secure quantum communication in quantum networks. QAN protocol has applications ranging from multiparty quantum computation to quantum internet. We utilize the QAN protocol to propose an anonymous quantum private comparison protocol in an $n$-node quantum network. This protocol can compare private information of any $2 \leq k \leq n$ parties with the help of the remaining $n-k$ parties and a semi-honest third party. These protocols feature a traceless property, i.e., encoding operations cannot be traced back to their originating sources. Security analysis shows that this protocol is robust against external adversaries and malicious participants.
\end{abstract}


\end{frontmatter}


\section{Introduction}
Development of quantum communication systems provides potential benefits to carry out the information processing tasks in a quantum network. Many quantum network based applications such as quantum secret sharing \cite{CGL:99:PRL, BMHMWRT:14:NC}, quantum voting \cite{BH:17:PRA}, and quantum conference key agreement \cite{RMW:18:PRA} have been proposed with the vision of quantum internet \cite{KIM:08:Nature}.

One practical application of quantum networks is quantum private comparison (QPC) protocols \cite{YW:09:JPA}. The main objective of QPC is to allow two parties to compare their private information with the help of a semi-honest third party (TP) (which can be almost dishonest \cite{ HHHK:17:QIP}), without leaking the information to participants and the TP. QPC has been generalized to compare secrets of multiple parties and is called multiparty QPC (MQPC) \cite{chang:13:QIP}. MQPC has many applications such as quantum voting \cite{JHNXZ:12:PRA}, quantum bidding \cite{MTKPZB:14:PRX}, and quantum auctions \cite{HHC:07:IJQI}. 

\begin{figure*}[t!]
	\centering
	\includegraphics[ width = 0.85 \textwidth]{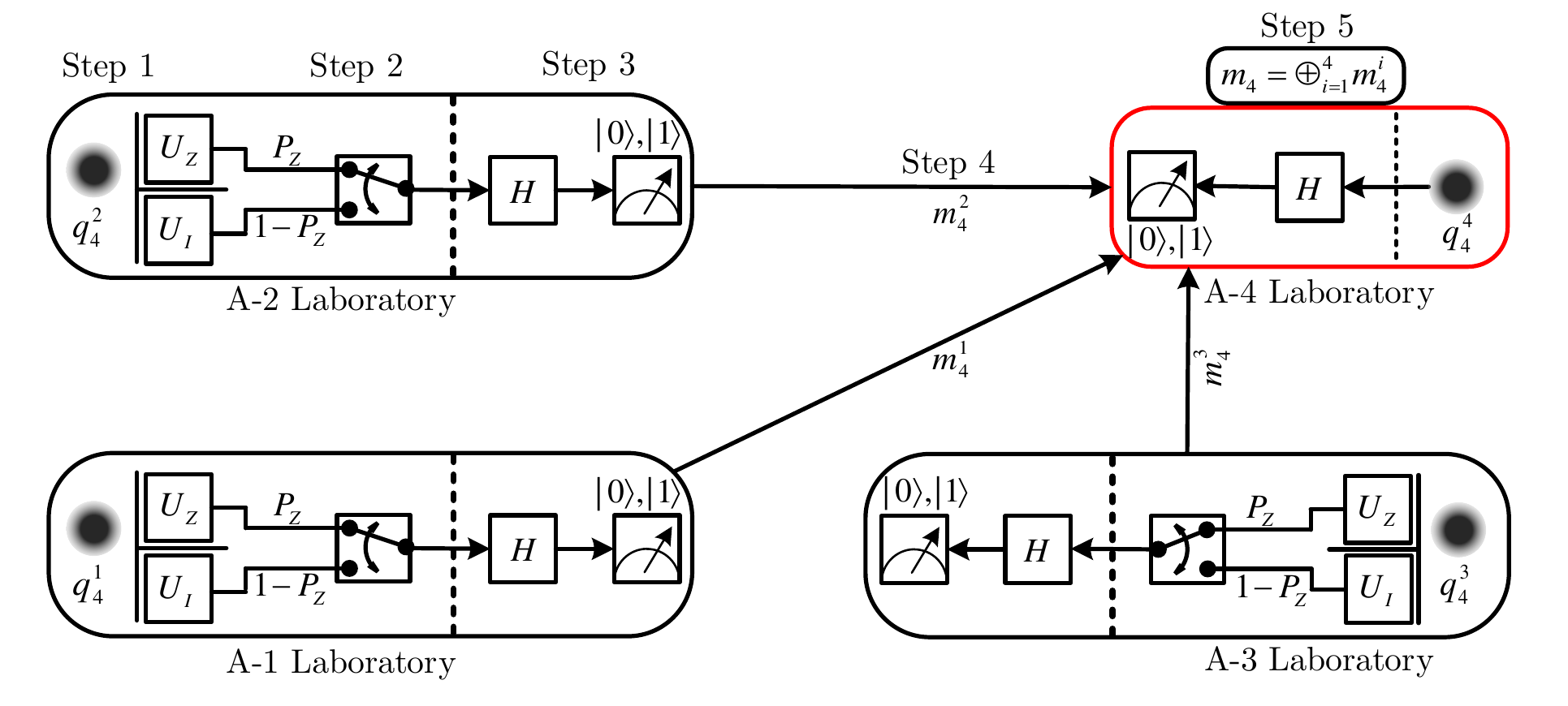}
	\caption{Anonymous quantum notification protocol.}
	\label{fig:1}
\end{figure*} 

One of the challenging requirements of quantum networks is to provide anonymity of the sender and the receiver when they wish to communicate through the network. Anonymity should be guaranteed without making any assumption on the computational power of malicious participants who might have quantum computers. The introduction of anonymity provides security against maliciously behaved participants and potential adversaries.
   
In the classical scenario, Broadbent \emph{et. al.} showed how to transmit a message anonymously with information-theoretic security in the absence of an honest majority \cite{BT:07:IC}.  The key enabler for the anonymity in their proposal is an anonymous notification protocol, which utilizes authenticated pairwise private channels and simultaneous classical broadcast channels.

In the case of quantum communication, the first work related to an anonymous message was proposed by Christandl and Wehner \cite{CW:05:IC}. They proposed the protocols with perfect anonymity for quantum broadcast and the creation of an EPR pair between two remote parties.  More recently, the protocol for anonymity in a practical quantum network was proposed \cite{UMYDMK:19:PRL}. The main ingredient of the protocol which introduces the anonymity in the network is a classical notification protocol. However, this classical notification protocol can be vulnerable to quantum computers. That means communicating parties are no longer anonymous against a sufficiently powerful quantum adversary.

We present the first quantum anonymous notification (QAN) protocol to introduce anonymity in a practical quantum network. This protocol guarantees the anonymity of communicating parties and also features traceless property, i.e., once the receiver is notified there is no way to trace the notifier. We exemplify the application of QAN on a new MQPC, where the QAN acts as the main ingredient for providing the anonymity.
Our MQPC protocol compares private information anonymously of any $2 \leq k \leq n$  parties with the help of the remaining $n-k$ parties and the semi-honest TP. In contrast to previous MQPC protocols, our proposed protocol removes the requirement of use of hash function, quantum key distribution, quantum secure direct communication, or multiple TPs \cite{YE:16:CTP, HHHK:17:QIP, AAHHFMG:19:IEEE_A, AFHMG:19:SR}.  We prove the security against common attacks launched by the participants, TP, and outsiders.

\section{Quantum Anonymous Notification}
Here we provide the QAN protocol for a quantum network where any of the participants can anonymously notify another participant for any upcoming targeted communication task.

\textbf{Communication scenario}.---Our network consists of $n$ agents that can perform local operations and classical communication (LOCC).  GHZ states are shared between these $n$ agents using the protocol in \cite{PCWDK:12:PRL, MPBMCLMMDKO:16:NC}. We also require pairwise classical authenticated channels between agents. The communication objective is to allow any honest party to anonymously notify any other party in the network. The adversaries and malicious agents aim is to break the anonymity or security of the protocol.

\noindent\rule{16.5cm}{0.8pt}

\textbf{Protocol 1} Quantum Anonymous Notification \\[-0.6cm]

\noindent\rule{16.5cm}{0.3pt}

\textbf{Prerequisite:} $n$ $(n)$-partite GHZ states.

\textbf{Protocol Parameters}\\[-0.5cm]
\begin{itemize}
\item Sender choice of the receiver is party $r$
\item 
$
U_{I}=
\begin{bmatrix}
1 & 0 \\
0 & 1 
\end{bmatrix},
$
 $\,
U_{Z}=
\begin{bmatrix}
1 & 0 \\
0 & -1 
\end{bmatrix}
$
 \end{itemize}
\textbf{The Protocol}
\begin{itemize}
\item[1.] 
GHZ states and the constituent particles of all GHZ states are numbered such that $q_{j}^{i}$ is the $i$th particle in the $j$th GHZ state. Each party $i$ is given $n$ particles,  $q^{i}_{j}, 1\leq j\leq n$. 

\item[2.]
Each party $i$ applies $U_j^i$ on the $q^{i}_{j}$ according to the rule:
\begin{align}
	U_{j}^i = \begin{cases}
			U_{Z}, \text{ with probability } P_Z,\\
			U_{I},  \text{ with probability } 1 - P_Z,
	\end{cases}
\end{align}
if $j = r$ is the intended party to be notified. Otherwise $U_{j}^i = U_I$ with probability 1.

   
%
%
\item[3.]
 Each party applies $H$ to all of the held particles and measures them in the computational basis. The measurement results on $q_j^i$ is $m_j^i \in \left\{ 0,1\right\}$.
\item[4.] Each party $i$ announces the measurement results $m_j^i \, \forall j \neq i$ on a classical authenticated channel. 

\item[5.] Each party calculates $m_j =  \oplus_{i=1}^{n} m_{j}^i$ for their allotted GHZ state.

\item[6.] Steps 1--5 are repeated $K$ times. Party $j$ is notified if ${m_j = 1}$ for any run of the protocol.
\end{itemize}
\vspace{-0.3cm}
\noindent\rule{16.5cm}{0.3pt}

Here we analyze the Protocol 1. We assume that $(n)$-partite GHZ states are shared between the network,
$$\ket{\text{GHZ}} = \frac{1}{\sqrt{2}} \left(\ket{0}^{ \otimes n} + \ket{1}^{\otimes n}   \right). $$
In step 1),  each party is allotted with a GHZ state. 2) The sender simply notifies by applying $U_{Z}$ on the receiver related particle of GHZ state with a probability of $P_{Z}$. This introduces the phase on the receiver GHZ state with a probability $P_Z$.  3) After the $H$ gate state changes into the superposition of all strings $E_{j}\in \left\{0, 1 \right\}^{n}$ with an odd number of $1$'s if $j=r$ else even number of $1$'s:
\begin{align*}
 &H^{\otimes n} \ket{\text{GHZ}} \\
 &= \frac{1}{\sqrt{2^{n+1}}} \sum_{E_{j} \in  \{0, 1\}^n} \left( 1 +  (-1)^{v \oplus |E_{j}|}  \right)\ket{E_{j}},
\end{align*}
where $|E_{j}|$ is the hamming weight of string $E_{j}$.  After step $4$ \& $5$, each party calculates $m_i$ for their allotted GHZ state. If the party is notified then she will get odd number of $1$'s and $v=1$  else she is not notified and $v=0$.  This result $m_j $ is only available to the party $i$. Our protocol gives the perfect anonymity to the receiver and the sender. Malicious participants cannot obtain the identity of the notifier and the notified party even if they collaborate as long as the total number of collaborating parties is less than $n-2$.

Fig.~\ref{fig:1} shows the QAN protocol for $n=4$ on $q_4^i$ where any of the parties can notify the fourth party $A-4$.

\begin{figure*}[t!]
	\centering
	\includegraphics[ width = 0.85 \textwidth]{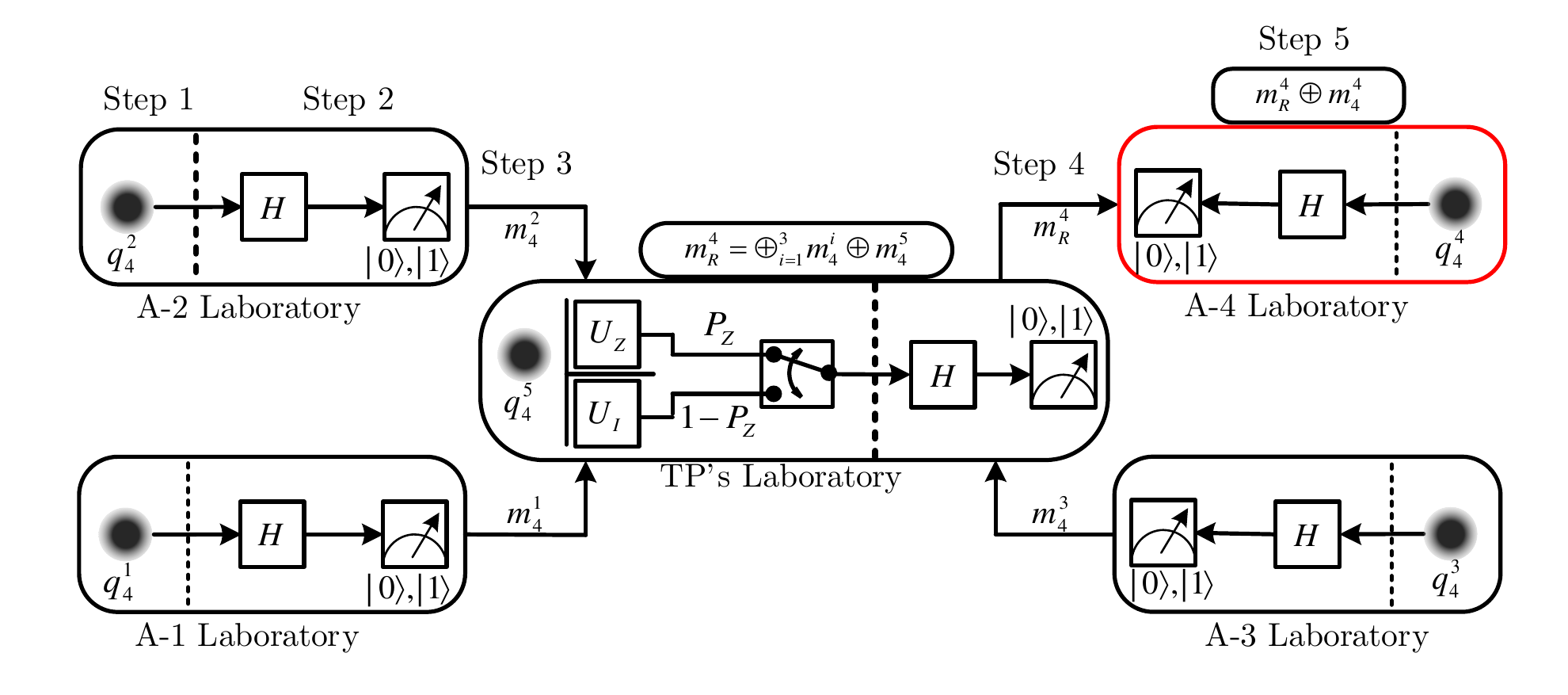}
	\caption{Modified anonymous quantum notification protocol.}
	\label{fig:2}
\end{figure*}

\section{Anonymous Multiparty Quantum Private Comparison}
In this section, we provide an application for quantum anonymity in multiparty computation protocol. We consider the problem of multiparty quantum private comparison. First, we provide the protocol for resource sharing, which will be used as a sub-protocol for the anonymous MQPC protocol. In this protocol, agents are connected via the quantum channel and classical authenticated channel.

\noindent\rule{16.5cm}{0.8pt}

\textbf{Protocol 2} Resource sharing and security check \\[-0.6cm]

\noindent\rule{16.5cm}{0.3pt}

\textbf{Protocol Parameters}\\[-0.7cm]
\begin{itemize}
	\item $n$: Total number of participants\\[-0.7cm]
	\item $L$: Total number of $(n)$-partite GHZ states shared at the end of the protocol\\[-0.7cm]
	\item $S$: Security parameter, the number of GHZ states utilized in the security check and verification.

\end{itemize}
\textbf{The Protocol}
\begin{itemize}
\item[1.]
\textbf{State Preparation:}
All parties in the network are indexed with $i\in \left\{ 1,2, \cdots, n\right\}$. A randomly designated player (e.g., Alice, with index $a$) prepares $K = L + S$ GHZ states and labels them as $F_j^i, \ i \in \left\{ 1,2, \cdots, n\right\}, \ j \in \left\{ 1,2, \cdots, K\right\}$. Here $F^i_j$ is the $i$th particle in the $j$the GHZ state. 
\item[2.] 
\textbf{State Distribution:} 
Alice sends $F_j^i, \ 1\leq j \leq K$ particles to $i$th party in the network through quantum channel and keeps $F_j^{a}$ for herself.  

\item[3.] 
\textbf{Security Check \& Verification: }
Participants repeat the following steps $S$ number of times:
\begin{enumerate}

\item[(i)] One of the party $i'\neq a$ randomly announces $j' \in \left\{ 1,2, \cdots , K \right\}$, and $h = 0,1$. Each party $i$ applies $H^h$ on $F^i_{j'}$ where $H$ is the Hadamard gate. 
\item[(ii)] Alice measures $F^{a}_{j'}$ in the computational basis and announces the outcome on the classical authenticated channel. After Alice's announcement, the rest of the parties perform the same measurement on their respective particles and announce their measurement results. 
\item[(iii)] The protocol aborts if 1) any party refuses to announce the measurement results, or 2) the measurement results are not consistent with the GHZ state.

\end{enumerate}
\item[4.] If the protocol does not abort, they have shared $L$ $(n)$-partite GHZ states with probability ${1 - 2^{-S}}$.
\end{itemize}
\vspace{-0.3cm}
\noindent\rule{16.5cm}{0.3pt}

After the resource sharing protocol, we move towards the first step of anonymous quantum private comparison. We use Protocol 1 to introduce the quantum anonymity in quantum private comparison network. In the quantum private comparison network,  TP performs the comparison between competing parties and is already known as the notifier.  Therefore, only the receiver's anonymity is required. Protocol 1 can be easily modified to provide receiver anonymity.  

In Protocol 1, small changes are made in steps 1-4.  1) $(n+1)$-partite GHZ states are shared instead of $(n)$-partite where the $(n+1)$th particle is held by the TP. 2) The TP only applies $U_{j}^{TP}$ on the $q_{j}^{TP}$  particle to notify the competing parties. 3) The TP also performs this step similarly on her held particle.  4) Each party announces the measurement result $m_j^i \, \forall j \neq i$ to TP via a classical authenticated channel.  TP calculates $m_{R}^{i} = \oplus_{i=1}^{n} m_j^i \oplus m_{j}^{TP}, \, \forall i \neq j $ for each party and sends it to the respective parties on a classical authenticated channel. After this procedure, if a party obtains $m_{j}=1$ for any run of the protocol then she is a competing party. Else she will assist the TP for the comparison. The simplified setup of this protocol is explained in Fig.~\ref{fig:2} for the four-party case. All parties and the TP already shared a five-partite GHZ state. The notification procedure is shown only for the A-4 party and similarly follows for other parties as well.

The comparison protocol starts after the successful run of the notification procedure. We first present the protocol for the two-party comparison in a network of $n$ parties.

\noindent\rule{16.5cm}{0.8pt}

\textbf{Protocol 3} Anonymous Quantum Private Comparison (For two parties in an $n$-partite network) \\[-0.6cm]

\noindent\rule{16.5cm}{0.3pt}

\textbf{Prerequisite:} From Protocol~2. $M$ $(n+1)$-partite and $M$ $(n)$-partite GHZ states whose particles are labelled as $G^i_j$ and $F^i_j$, respectively. $G_j^{n+1}, {1\leq j \leq M}$ are held by the TP.

\textbf{Protocol Parameters}\\[-0.7cm]
\begin{itemize}
	\item $M$: Total number of secret bits of each party  \\[-0.7cm]
	\item $n$: Total number of parties in the network\\[-0.7cm]
\end{itemize}

\textbf{The Protocol}
\begin{itemize}

\item[1.]
Each notified party (e.g., Alice and Bob, labelled with indexes $a$ and $b$, respectively) encode their secrets as follows.
\\For $1\leq j \leq M$:
\begin{enumerate}
\item[(i)]  Alice and Bob measure $F^a_j$ and $F^b_j$, respectively, in the computational basis. Let $k_j^a$ and $k_j^b$ be their measurement outcomes.
\item[(ii)] If $k_j^a = k_j^b = 0$, Alice and Bob encode their secret messages by applying $U_I(U_{Z})$ for the message bit 0(1) on $G^a_j$ and $G^b_j$, respectively. Else if $k_j^a = k_j^b = 1$, they reverse the role of $U_I$ and $U_Z$.
\end{enumerate}
\item[2.] After the encoding, each participant including Alice, Bob, and TP apply $H$  to their $G^i_j,$ qubits.
\item[3.] Each participant and TP measure their $G^i_j$ qubits in the computational basis and send their measurement outcomes $m^{i}_{j}$ to the TP via the classical authenticated channel.
\item[4.] The TP calculates the  $m_j = \oplus_{i=1}^{n+1} m^{i}_j$ for each secret bit $j$. The secrets of Alice and Bob are equal if $m_j = 0, 1\leq j \leq M$, otherwise they are not equal. The TP announces whether or not the secrets are equal. 
\end{itemize}
\vspace{-0.3cm}
\noindent\rule{16.5cm}{0.3pt}
\begin{figure*}[t!]
\centering
\includegraphics[ width = 0.85 \textwidth]{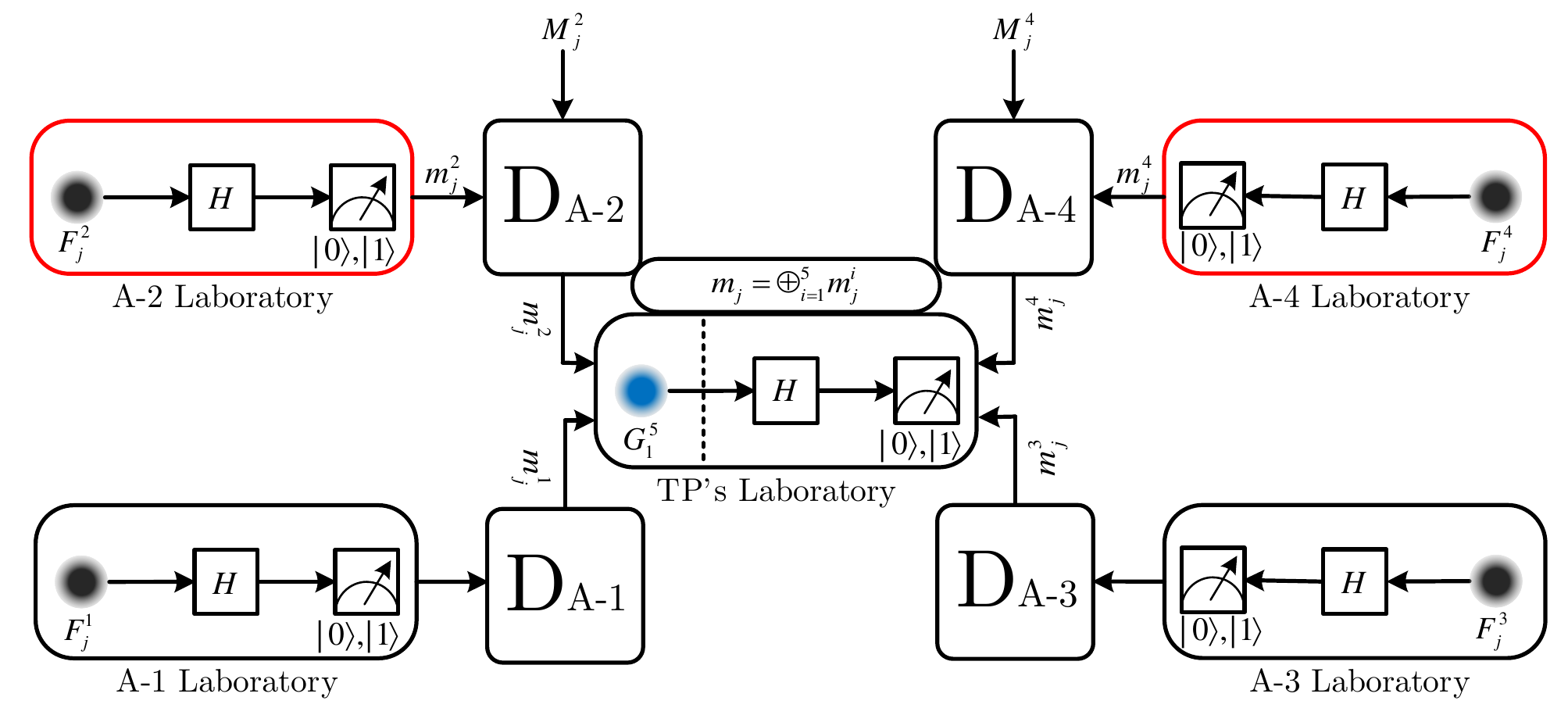}
\caption{Example run of anonymous quantum private comparison protocol for two comparing parties in a four-node network.}
\label{fig:3}
\end{figure*}

This protocol is also based on the LOCC only.  For example, the notified competing participants are Alice and Bob have a secret bit $b_{1}$ and $b_{2}$, respectively, where $b_{1}, b_2 \in \{0, 1 \}$. 

\begin{figure}[t!]
\centering
\includegraphics[ width = 0.45 \textwidth]{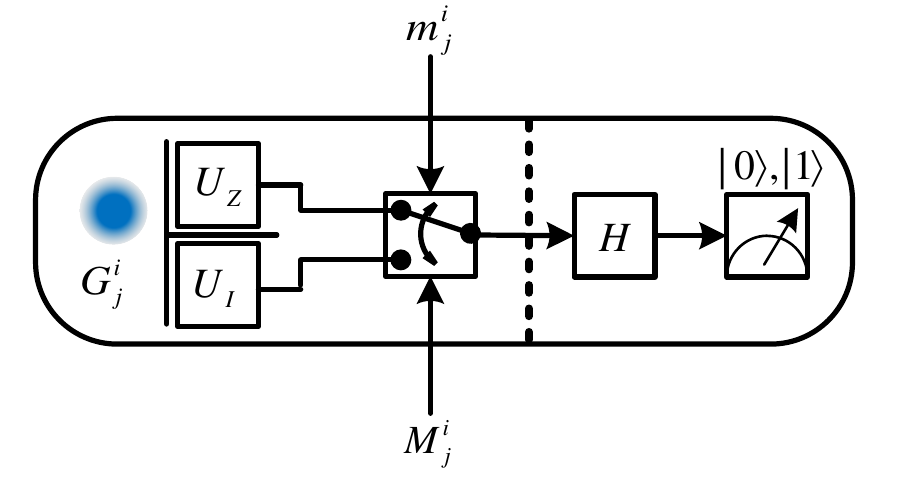}
\caption{A-i Device (D$_\text{A-i}$)  where $i \in \{1, 2, 3, 4\}$. $M_{j}^{i}$ represents jth secret bit of ith party and $M$ is the total number of secret bit.}
\label{fig:4}
\end{figure}

In step 1, Alice and Bob apply the unitaries $U_I$ or $U_Z$ on the particles of $(n+1)$-partite GHZ state for the message bit $b_{1}$ and $b_{2}$. This encoding depends upon the measurement outcome of their $(n)$-partite GHZ state particle.  After the encoding, the change of phase in global GHZ state depends upon the secret bits,
\begin{align*}
\ket{\text{GHZ}} = \frac{1}{\sqrt{2}}\left( \ket{0}^{\otimes n+1} + (-1)^{b_1 \oplus b_2} \ket{1}^{\otimes n+1} \right).
\end{align*} 
If the secret bits $b_1 \neq b_{2}$, then the phase of the global GHZ state changes else it remains the same.  Each participant and the TP applies the $H$ gate to their particles in step 2.  This changes the GHZ state into the superposition of all strings $E_{j} \in \{ 0, 1 \}$ with an even number of 1's for $b_1 = b_2$ or an odd number of 1's for $b_1 \neq b_2$. After $H$, state becomes:
\begin{align*}
  &H^{\otimes n+1}\ket{\text{GHZ}} \\ 
 &= \frac{1}{\sqrt{2^{n+2}}} \sum_{E_{j} \in  \{0, 1\}^{n+1}} \left(1 +   (-1)^{b_1 \oplus b_2 \oplus |E_{j}|}  \right)\ket{E_{j}},
\end{align*} 
where $|E_{j}|$ is the hamming weight of string $E_j$.  TP computes the phase of the GHZ state after step 3. If the result is $m_{j} =0$ for each secret bit $j$ then secret is equal else secrets are not equal. In the Fig.~\ref{fig:3} \& \ref{fig:4} simplified example of the experimental setup for a four-partite case is shown. Fig.~\ref{fig:4} explains the inside of the device of participants. Fig.~\ref{fig:3} explains that A-2 and A-4 laboratories are comparing secrets anonymously with the help of the remaining parties and the TP.   


In the following, we present the protocol of private comparison between more than two parties. This protocol is similar to the Protocol 3 with a slight modification.  In this protocol, both the unitaries are utilized for a single secret bit.

\noindent\rule{16.5cm}{0.8pt}

\textbf{Protocol 4} Anonymous Quantum Private Comparison (For $n > 2$) \\[-0.6cm]

\noindent\rule{16.5cm}{0.3pt}

\textbf{Prerequisite:}\\[-0.7cm]
\begin{itemize}
	\item $K$ $(n+1)$-partite GHZ states, where $K = \left(2\lceil\log( n ) \rceil + 2 \right)M$. The particle held by the $i$th party, which will be utilized for comparing the $m$th bit of the message is labelled as $Q^{m, i}_{k,g}$. The role of $k \in \left\{ 0,1\right\}$, and $g\in \left\{0, 1, \cdots \left\lceil\log(n)\right\rceil\right\}$ will be explained in the protocol. $Q^{m, n+1}_{k,g}, \forall (m, k, g)$ are held by the TP.\\[-0.7cm]
	 \item $M$ $n$-partite GHZ states. The particle held of the $j$th state held by the $i$th party is labeled as $F^i_j$.
\end{itemize}
\textbf{Protocol Parameters}\\[-0.7cm]
\begin{itemize}
	\item $M$: Total number of secret bits of each party\\[-0.7cm]
	\item $n$: Total number of parties in the network\\[-0.7cm]
	\item  $ U_{g} (\phi) = \cos(\phi/2) U_I + i\sin(\phi/2)U_Z$, where $\phi = \pi/2^{g}$ and $0 \leq g \leq \lceil \log  (n) \rceil$.
\end{itemize}
\textbf{The Protocol}
\begin{itemize}

\item[1.]
Each notified party (e.g. Alice, Bob, and Charlie, labelled with indexes $a$, $b$, and $c$, respectively) encode their secrets as follows.

Steps (i)--(ii) are repeated for $1\leq m \leq M$:

\begin{enumerate}

\item[(i)]  Alice, Bob, and Charlie measure $F^a_m$, $F^b_m$ and $F^c_m$, respectively, in the computational basis. Let $k_m^a = k_m^b = k_m^c = k$ be their measurement outcomes.
\end{enumerate}

Step (ii) is repeated for $0 \leq g \leq \log(n)$:
\begin{enumerate}
\item[(ii)] Alice, Bob,  and Charlie encode their secret bits by applying $U_I(U_{g})$ for the message bit 0(1) on $Q^{m,a}_{k,g}$, $Q^{m,b}_{k,g}$ and $Q^{m,c}_{k,g}$, respectively. Then, they reverse the role of $U_I$ and $U_g$ and perform the encoding on $Q^{m,a}_{\neg k,g}$, $Q^{m,b}_{\neg k,g}$ and $Q^{m,c}_{\neg k,g}$, where $\neg$ denotes the logical NOT.
\end{enumerate}

\item[2.] All participants including the TP apply $H$ to their $Q^{m,i}_{k,g}$ qubits.

\item[3.]  All participants including the TP measure their $Q^{m,i}_{k,g}$ qubits in the computational basis and send the respective measurement outcomes $z^{m,i}_{k,g}$ to the TP via the classical authenticated channel.

\item[4.] For each message bit  $1\leq m \leq M$, the TP calculates $ \mathcal{D}^{m}_{k,g} = \oplus_{i=1}^{n+1} z^{m,i}_{k,g}, \forall (k,g) $. The $m$th message bit of all participants are equal if $\mathcal{D}^{m}_{k,g} = 0, \forall g$ when $k$ is fixed to either 0 or 1. The TP announces whether the secrets are equal or not. 
\end{itemize}
\vspace{-0.3cm}
\noindent\rule{16.5cm}{0.3pt}
The correctness of this protocol can be deduced from the correctness of Protocol 3.

 \section{Security Analysis}
The security of quantum private comparison protocols is hard to prove due to the participant's and the TP's attacks. Previously, these protocols were vulnerable to participants' attacks.  However,  the secrecy of protocol increases due to the induction of anonymity and traceless properties.  For security, we analyze the quantum communication between the participants and the TP.
\begin{itemize}
\item \textbf{Outside attacks :}
A designated player (e.g., Alice) shared the GHZ states via the quantum channel with the parties and TP. After receiving the GHZ states, one party randomly selects GHZ state for security check and also announces a random bit $h$. The value of $h$ determines that all parties and TP should perform Hadamard transformation or not. This process not only detects the adversaries but also inquires about Alice's honesty, whether she prepared the GHZ states correctly. After the selection of GHZ state and value of $h$, Alice measures her particle in the computational basis and announces the results. Then, each party and the TP measure their qubits and announce the result to check the security of the communication. The attacker does not know in advance the value of $h$ and the GHZ state chosen by the random party. This process will prevent well-known attacks such as intercept-resend attacks, entangled-resend attacks, and correlation elicitation attacks. The detection of the adversary is $P_d=1-2^{-S}$, where $S$ is the security parameter.  

\item \textbf{Participants attacks :}
The TP anonymously notifies the competing participants.  To get the secret information, the participants have to identify first the competing parties. To reveal the identity of competitors is impossible for a single participant.  After the encoding, the only communication between participants and the TP is via the classical authenticated channel. This information does not reveal any information to the participants. This communication only helps the TP to compute the comparison results. Therefore, cheating and collusion attacks are not possible for the proposed protocol. The private information is still not available for the adversaries if the $n-1$ parties collude against the one.  The only option left for malicious parties is to adopt Eve's strategies to steal private information.  These strategies are known as outside attacks. As discussed earlier, the protocol is secure against these attacks.
  
\item \textbf{TP's attacks :}
Malicious TPs can also threaten the security of the private comparison protocols. The TP can attack to gain useful information using her resources.  Our procedure has a unique traceless property to tackle the TP's attacks. Firstly, with the assumption that semi-honest TP is not allowed to collude with the participants. To prove the security, assume that competing parties (e.g., Alice and Bob) have secrets of 0 and 1, respectively. After applying the unitaries, the GHZ state has a phase, which means that the secrets are not equal. The TP computes the phase of the GHZ state with her and each participant's measurement outcomes. The TP cannot identify the party which introduced the phase. So this protocol is secure against the TP's attack as well.

\end{itemize}

\section{Conclusion}\label{sec:conc}
Anonymity can be a useful but challenging requirement of any communication network.  In this work, we proposed the QAN protocol which can be implemented in quantum networks where anonymity is required (e.g., accessing the quantum internet without revealing the identity). QAN provides both sender and receiver anonymity. However, this protocol can be easily modified to provide receiver anonymity only. 
We used this protocol as an application and introduced the anonymity in the MQPC.  Firstly, TP notifies the participants via the QAN protocols. Then, the MQPC protocol is executed to anonymously and securely perform the comparison of secrets of different remote parties. Our framework allows two or more parties to compare their secrets with the help of the remaining parties and a semi-honest TP.  Security analysis shows that this protocol is robust against malicious participants and the TP.

\section*{Conflict of interest}
The authors declare that they have no conflict of interest.



\section*{Acknowledgments}
This work was supported by the National Research Foundation of Korea (NRF) grant funded by the Korea government (MSIT) (No. 2019R1A2C2007037).

\section*{Author contributions}

A.K contributed the idea. A.K and J.R developed the theory and wrote the manuscript. H.S improved the manuscript and supervised the research.  All the authors contributed in analyzing and discussing the results and improving the manuscript.


\end{document}